\documentclass[preprint,showpacs,amsmath,amssymb]{revtex4-2}

\usepackage{amsmath}
\usepackage{graphicx}
\usepackage{amsfonts}
\usepackage{dcolumn}
\usepackage{bm}
\usepackage{color}

\newcommand \la{\langle}
\newcommand \ra{\rangle}

\begin{document}

\title{Dynamic and Thermodynamic Origins of Motility-Induced Phase Separation}

\author{Jie Su$^{1,2}$}
\thanks{These authors have contributed equally to this work.}
\author{Zhiyu Cao$^{1}$}
\thanks{These authors have contributed equally to this work.}
\author{Jin Wang$^{2,3}$}
\thanks{E-mail: jin.wang.1@stonybrook.edu}
\author{Huijun Jiang$^{1}$}
\thanks{E-mail: hjjiang3@ustc.edu.cn}
\author{Zhonghuai Hou$^{1}$}
\thanks{E-mail: hzhlj@ustc.edu.cn}

\affiliation{1.Department of Chemical Physics \& Hefei National Research Center for Physical Sciences at the Microscale, University of Science and Technology of China, Hefei, Anhui 230026, China}
\affiliation{2.Center for Theoretical Interdisciplinary Sciences, Wenzhou Institute, University of Chinese Academy of Sciences, Wenzhou 325001, China}
\affiliation{3.Department of Chemistry and of Physics and Astronomy, State University of New York of Stony Brook, Stony Brook, New York 11794, USA}
\date{\today}

\begin{abstract}
Active matter systems are inherently out of equilibrium and break the detailed balance (DB) at the microscopic scale, exhibiting vital collective phenomena such as motility-induced phase separation (MIPS). Here, we introduce a coarse-grained mapping method to probe DB breaking in the density-energy phase space, which allows us to reveal the dynamic and thermodynamic origins of MIPS based on nonequilibrium potential and flux landscape theory. Hallmarks of nonequilibrium properties are manifested by identifying the visible probability flux in the coarse-grained phase space. Remarkably, the flux for the system with the activity lower than the MIPS threshold tends to ``tear up" the single potential well of the uniform-density phase to create two wells of phases with different densities, presenting directly that the nonequilibrium flux is the dynamic origin of MIPS. Moreover, we find that the obtained entropy production rate (EPR) of the system undergoes a transition from nearly independent of activity to increasing proportionally as activity increases after the single well is "teared up". The transition of EPR's scaling behavior might provide a hint of the thermodynamic origin of MIPS in the coarse-grained space. Our findings propose a new route to explore the nonequilibrium nature of active systems, and provide new insights into dynamic and thermodynamic properties of MIPS.
\end{abstract}

\maketitle
Active matter widely exists in various scales of nature, ranging from microscopic and mesoscopic swimmers such as bacteria and the active Janus spheres, to macroscopic objects, such as fish, birds and horses \cite{a1,a2}. Since the active systems break the detailed balance (DB) at the microscopic scale, they cannot be described by equilibrium statistical mechanics, and the nonequilibrium dynamics may manifest as curl flux in a phase space of mesoscopic coordinates \cite{B1,B2,B3,B4,B5,B6,B7,B8}. In comparison with their passive counterparts, active systems exhibit many novel dynamical behaviors, such as the emergence of dynamic chirality \cite{a4,a5,a6,a7}, functional self-assembly \cite{a8,a9,a10,a11}, abundant collective motions such as vortex, swarm \cite{a11,a12,a13,a14,a15}, and particularly well-known motility-induced phase separation (MIPS) \cite{MIPS1}. It has been observed that, particles with pure repulsion can spontaneously undergo phase separation between dense and dilute fluid phases when the activity is higher than a certain threshold. The unique nonequilibrium properties of MIPS have attracted tremendous research interests \cite{MIPS1,MIPS2,MIPS3,MIPS4,MIPS5,MIPS11,MIPS12,MIPS13,MIPS14,MIPS15,MIPS16,MIPS17,MIPS18,MIPS19,MIPS20,MIPS21,MIPS22,MIPS23,MIPS24,MIPS25,MIPS26,MIPS9,MIPS10,MIPS6,MIPS7,MIPS8}.

It is well-known that some physical properties of equilibrium systems have singularities at the transition point, such as the discontinuous enthalpy change in the first-order transition or the discontinuous change in heat capacity in the second order transition\cite{landau2013statistical}. Nevertheless, it is quite difficult to define state functions of nonequilibrium systems, such as Gibbs free energy, so it is very important but challenging to figure out the underlying mechanism of MIPS by other means. Previously, the ``self-trapping" effect of active particles was proposed by Cates et al to understand the mechanism of MIPS \cite{MIPS1}. That is, active particles tend to accumulate where they move more slowly and will slow down at high density, which then creates positive feedback and further leads to MIPS \cite{MIPS1,MIPS2}. On this basis, a series of theoretical analysis, including the kinetics model \cite{MIPS3,MIPS4,MIPS5}, swim pressure \cite{MIPS11,MIPS12,MIPS13}, scalar $\phi^4$ field theory \cite{MIPS9,MIPS10} and effective Cahn-Hilliard equation \cite{MIPS6,MIPS7,MIPS8}, are performed to explore the deep insight of MIPS. However, the dynamic and thermodynamic origins of the occurrence of MIPS and their connection to how DB violation propagates from the particle-scale dynamics to the large-scale collective dynamics in active matter systems remain unclear.

In this letter, we introduce a constructive method mapping the real space including motions of active particles into a low-dimensional phase space with two dimensions of local particle density and local particle energy. In the coarse-grained phase space, the theoretical approach of nonequilibrium potential and flux landscape theory \cite{flux1,flux2,flux3,flux4,flux5,flux7} is applied to explore the dynamic and thermodynamic origins of MIPS. The obtained nonequilibrium potential has only one potential well (representing the single phase) for activity before the MIPS threshold. Interestingly, we find that the flux inside the well tends to push local states of low density to be increasingly lower, and those of higher density to be increasingly higher. In other words, the flux tries to ``tear up" the potential well to create new wells of different densities. Further analysis reveals that the contribution of the nonequilibrium flux depends nonmonotonically on particle activity and shows a maximum value at the MIPS threshold, demonstrating that the nonequilibrium flux is the dynamic origin of the occurrence of MIPS. Moreover, it is observed that the obtained entropy production rate (EPR) is nearly unchanged before the threshold, and increases rapidly as activity increases after the threshold. The transition of the scaling behavior of EPR on activity might be considered as the thermodynamic origin of MIPS.

\emph{Coarse-grained mapping method}.--We propose a constructive method mapping the real space including motions of active particles into a low-dimensional coarse-grained phase space with two dimensions of local particle density and local particle energy as follows. For $N$ spherical active Brownian particles (ABPs) of diameter $\sigma$ and friction coefficient $\gamma$ in a quasi two-dimensional space with size $L$ and periodic boundary conditions, the motion of the \emph{i-th} ABP located at $\mathbf{r}_{i}$ obeys the following overdamped Langevin equations,

\begin{equation}
\mathbf{\dot{r}}_i=v_0\mathbf{n}_i-\gamma^{-1}\sum^N_{j=1,j\neq i}\nabla_{\mathbf{r}_i}U({r}_{ij})+\boldsymbol{\xi}_i,\label{eq:translation1}
\end{equation}
\begin{equation}
\dot{\theta}_i=\zeta_i.\label{eq:rotation1}
\end{equation}

\noindent Herein, $v_0$ and $\mathbf{n}_i=(\cos(\theta_i),\sin(\theta_i))$ are the amplitude and direction of active speed, respectively, with $\theta_i$ being the angle of $\mathbf{n}_i$. $\mathbf{r}_{ij}=\mathbf{r}_i-\mathbf{r}_j$ is the vector pointing from the \emph{j-th} ABP to the \emph{i-th} ABP, and $r_{ij}$ is its norm. The interaction between a pair of ABPs is described by the purely repulsive Weeks-Chandler-Andersen (WCA) potential, $U({r}_{ij})=4\epsilon[(\sigma/r_{ij})^{12}-(\sigma/r_{ij})^6+1/4]$ for $r_{ij}<2^{1/6}\sigma$, and $U({r}_{ij})=0$ otherwise, where $\epsilon$ is the interaction strength. $\boldsymbol{\xi}_i$ and $\zeta_i$ denote the independent Gaussian white noises with time correlations $\la\boldsymbol{\xi}_i(t)\boldsymbol{\xi}_j(t^{\prime})\ra=2D_t\mathbf{I} \delta_{ij}\delta(t-t^{\prime})$ and $\la\zeta_i(t)\zeta_j(t^{\prime})\ra=2D_r\delta_{ij}\delta(t-t^{\prime})$, where $\mathbf{I}$ is the unit matrix, $D_t=k_BT/\gamma$ is the translation diffusion coefficient with $k_B$ the Boltzmann constant and $T$ the temperature, and $D_r$ is the rotational diffusion coefficient which couples with the translational diffusivity $D_r=3D_t/\sigma^2$.

\begin{figure}
\centering
\includegraphics[width=1.0\columnwidth]{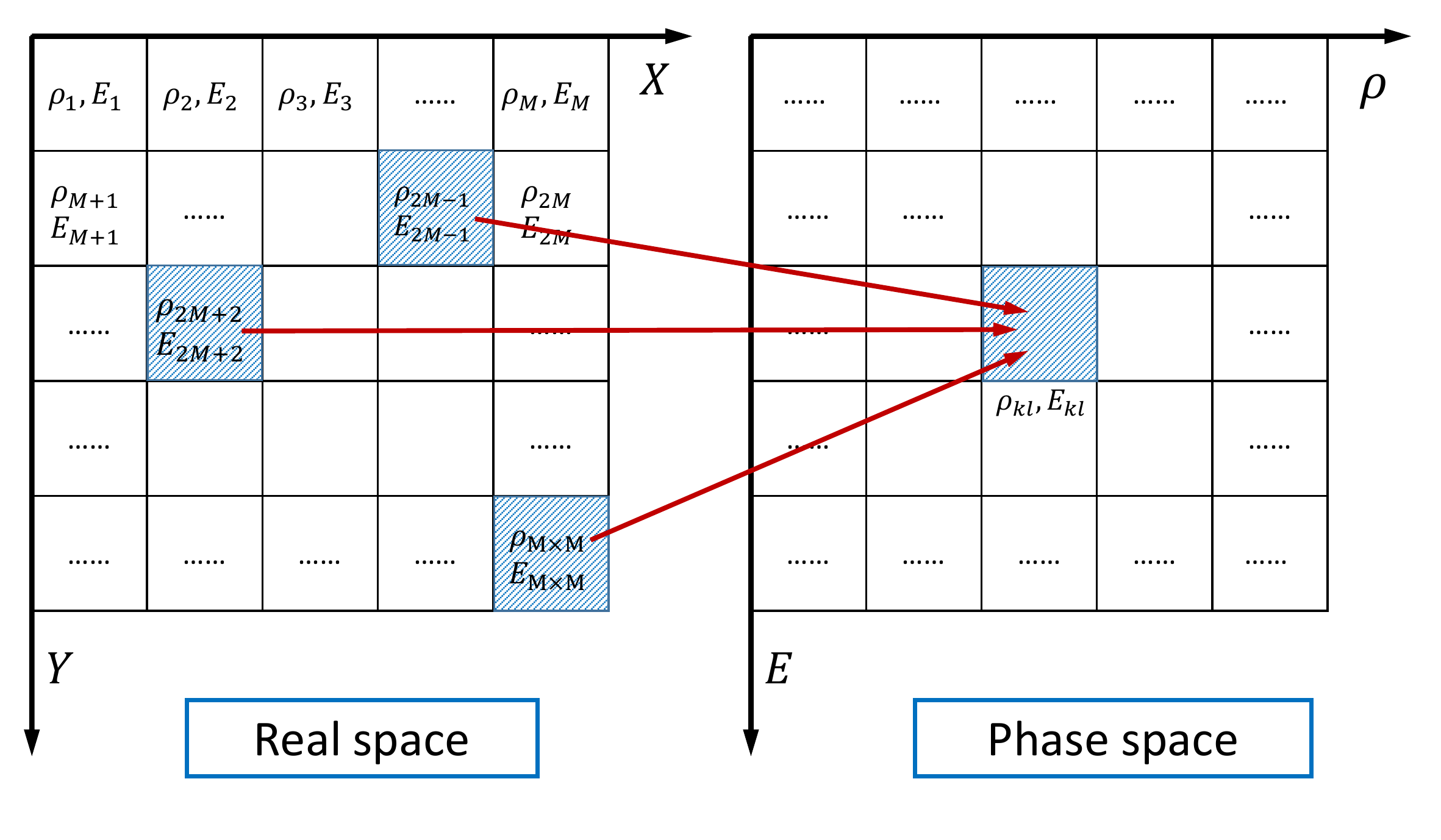}
\caption{Schematic of the coarse-grained mapping from the evolution of cells in real space to that in the $\rho-E$ phase space. As an example, if the three blue cells located at different positions in the real space have the same $\rho$ and $E$, they will be mapped into the same cell with $\rho_{kl}=\rho$, $E_{kl}=E$ in the phase space.}
\label{fig:result1}
\end{figure}

Based on the motion equations Eq.~\eqref{eq:translation1} and \eqref{eq:rotation1}, we divide the real space into $M\times M$ cells with size $a\times a$  ($M=L/a$) as shown in the left panel in Fig. \ref{fig:result1}. In the \emph{i-th} cell, two coarse-grained variables, local particle density and local total energy, can then be calculated as $\rho_i(t)=n_i/a^2$ where $n_i$ is the number of particles in the \emph{i-th} cell and $E_i(t)=\sum_{q\in cell_i}v_q^2(t)/2$. Similarly, the density-energy ($\rho-E$) space can also be divided into cells with steps $\delta\rho$ along the $\rho$ dimension and $\delta E$ along the $E$ dimension (the right panel in Fig. \ref{fig:result1}). Then, the \emph{i-th} cell in real space can be mapped into the $(k, l)$ cell in $\rho-E$ space if $\rho_i(t)\in[k\delta\rho,(k+1)\delta\rho)$ and $E_i(t)\in[l\delta E,(l+1)\delta E)$ at time $t$. If there are different cells in real space of the same $\rho(t)$ and $E(t)$, they will be mapped into the same cell in $\rho-E$ space (blue cells in Fig. \ref{fig:result1}). Hence, the probability distribution $P(\rho,E,t)$ in the $\rho-E$ phase space can then be calculated. For convenience, variables $\rho$ and $E$ with one subscript denote the local density and local energy of cells in real space, respectively, and those with two subscripts are those of cells in $\rho-E$ space in follows, if not otherwise stated.

As will be proven later, the density-energy phase space is sufficiently coarse-grained, where the non-Markovity of active matter systems does not manifest at such scale. Hence, the nonequilibrium dynamics of each cell in the $\rho-E$ phase space can be written as:

\begin{equation}
\dot{\rho}_{kl}(t)=F_\rho(\rho_{kl},E_{kl})+\xi_\rho(\rho_{kl},E_{kl},t).\label{eq:evolution-rho}
\end{equation}
\begin{equation}
\dot{E}_{kl}(t)=F_E(\rho_{kl},E_{kl})+\xi_E(\rho_{kl},E_{kl},t).\label{eq:evolution-E}
\end{equation}

\noindent Herein, $\dot{\rho}_{kl}(t)$ and $\dot{E}_{kl}(t)$ are calculated by averaging over mapping cells in real space. $F_\rho$ and $F_E$ are the determinate ``driving force" with $F_\rho(\rho_{kl},E_{kl})=\la\dot{\rho}_{kl}(t)\ra$ and $F_E(\rho_{kl},E_{kl})=\la\dot{E}_{kl}(t)\ra$ ($\la\cdot\ra$ denotes time averaging when the system reaches the steady state). $\xi_\rho$ and $\xi_E$ are the stochastic terms with time correlations $\la\boldsymbol{\xi}(\rho_{kl},E_{kl},t)\boldsymbol{\xi}(\rho_{kl},E_{kl},t^\prime)\ra=2\mathbf{D}(\rho_{kl},E_{kl})\delta(t-t^\prime)$, where $\boldsymbol{\xi}$ is a vector including the components of $\xi_\rho$ and $\xi_E$, and $\mathbf{D}(\rho_{kl},E_{kl})$ is a $2\times2$ diffusion coefficient tensor consisting of $D_{\rho\rho}$, $D_{\rho E}$, $D_{E\rho}$ and $D_{EE}$. This diffusion coefficient tensor can be obtained by the Fourier transform of the time correlation functions of the stochastic term $\boldsymbol{\xi}$. More details can be found in the Supplementary Information (SI). The power spectrum functions for $v_0=100$ are shown in Fig.S1 in SI as an example, which proves that $\xi_\rho$ and $\xi_E$ are white noises, demonstrating that the dynamics of the active system in the $\rho-E$ phase space is indeed Markovian.

\emph{Nonequilibrium potential and flux in the density-energy space}.--Based on our coarse-grained mapping method, both the probability distribution $P(\rho,E,t)$ and the dynamical equation of cells [Eqs.~\eqref{eq:evolution-rho} and \eqref{eq:evolution-E}] in the $\rho-E$ phase space can be obtained from the numerical simulations of active systems. We then apply the nonequilibrium potential and flux theory \cite{flux1,flux2,flux3,flux4,flux5,flux7} to figure out the dynamic and thermodynamic origins of MIPS in the density-energy phase space.

When the active system reaches the steady state, the probability distribution $P(\rho,E,t)$  is unchanged over time, i.e., $\partial P_{ss}(\rho,E,t)/\partial t=0$ (the subscript $ss$ denotes the steady state). The effective nonequilibrium potential $U_{neq}$ can then be defined naturally as $U_{neq}(\rho,E)=-\ln P_{ss}(\rho,E)$. According to the Fokker-Planck equation, it is known that $\partial P/\partial t=-\nabla\cdot\mathbf{J}(t)$. When the system is in equilibrium, the flux $\mathbf{J}=0$ so that the driving force only depends on the gradient of the effective potential, i.e., $\mathbf{F}=(F_\rho,F_E)^{\rm T}=-\mathbf{D}\cdot\nabla U$ (the superscript ${\rm T}$ denotes the transpose of a matrix). However, when the system is in nonequilibrium, the flux $\mathbf{J}$ can exist in the form of a rotational curl or more precisely recurrent field \cite{flux1,flux2,flux3,flux4,flux5,flux7}, and the steady-state nonequilibrium flux $\mathbf{J}_{ss}(\mathbf{x})$ with $\mathbf{x}=(\rho,E)^{\rm T}$ can be described as $\mathbf{J}_{ss}(\mathbf{x})=\mathbf{F}(\mathbf{x})P_{ss}(\mathbf{x})-\nabla_{\mathbf{x}}\cdot [\mathbf{D}P_{ss}(\mathbf{x})]$. Here, the driving force $\mathbf{F}$ no longer depends solely on the gradient of the nonequilibrium potential $U_{neq}$, because the steady-state flux field $\mathbf{J}_{ss}$ contributes to $\mathbf{F}$. Therefore, $\mathbf{F}$ can be decomposed into the gradient part ($\mathbf{F}_{gradient}$), the curl one ($\mathbf{F}_{curl}$) and the one related to the spatial dependent noise ($\mathbf{F}_{D}$), $\mathbf{F}(\mathbf{x}) =\mathbf{F}_{gradient}(\mathbf{x})+\mathbf{F}_{curl}(\mathbf{x})+\mathbf{F}_{D}(\mathbf{x})=-\mathbf{D}\cdot\nabla U_{neq}(\mathbf{x})+\mathbf{v}_{ss}(\mathbf{x})+\nabla\cdot\mathbf{D}$ (in Ito's representation, $\nabla\cdot\mathbf{D}$ represents the divergence of the diffusion tensor \cite{flux1}). Here $\mathbf{v}_{ss}(\mathbf{x})=\mathbf{J}_{ss}(\mathbf{x})/P_{ss}(\mathbf{x})$ is the local flow or velocity in steady states \cite{seifert2012stochastic}.

Combining our coarse-grained mapping method with the nonequilibrium potential and flux theory, it is convenient to analyze the dynamic and thermodynamic properties of MIPS in density-energy phase space based on numerical simulations in real space. In simulations, $\sigma$ and $k_BT$ are basic units for length and energy respectively, and $\gamma=1.0$ so that the basic unit for time is $\gamma\sigma^2/(k_BT)$. We fix $L=200$, $\epsilon=1.0$, and $N=30720$ so that the averaged number density becomes $\rho_0=N/L^2=0.768$ and the volume fraction is $\phi_0=\pi\sigma^2\rho_0/4=0.6$. Active systems with such volume fraction will undergo phase transition from the single phase to the coexisting phase at a threshold around $v_0=55$. We first run simulations from a random initial configuration for a long time $t_1=500$ with the time step $\Delta t=10^{-5}$ to ensure that the system reaches the steady state, and then perform the simulations for another long time $t_2=500$ to attain the steady-state evolution of ABPs. In the following, the data of local density and local energy are rescaled by $\rho_0$ and the averaged local energy $E_0$, i.e., $\rho^{\prime}=\rho/\rho_0$, $E^{\prime}=E/E_0$.

\begin{figure}
\centering
\includegraphics[width=1.0\columnwidth]{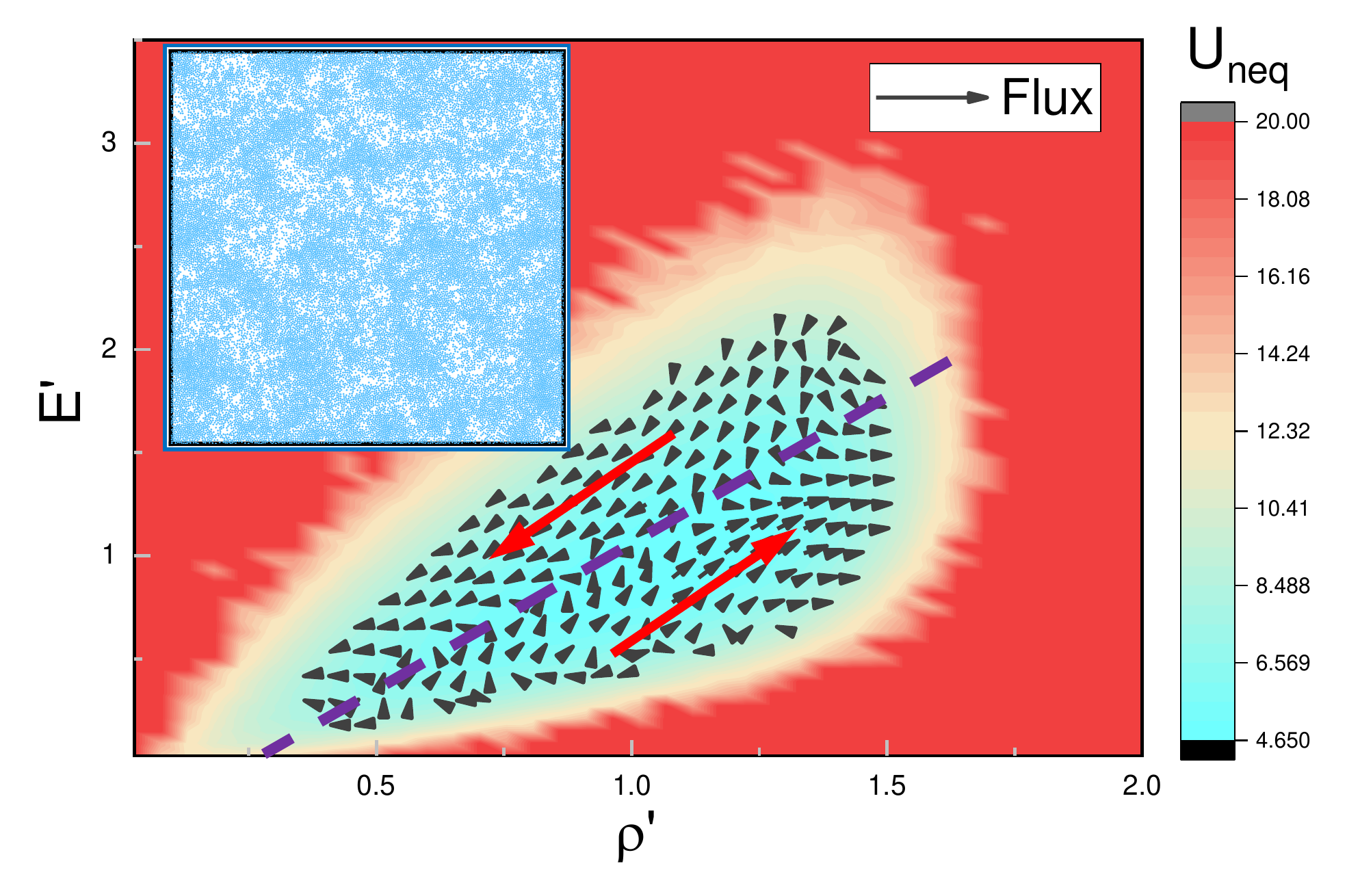}
\caption{The nonequilibrium potential and flux landscape of the active system with $v_0=40$. The nonequilibrium potential is presented as the colored background and the flux field is shown as the black arrows. The purple line is the division line of the flux direction and the red arrows represent the whole flux direction of each segment. The inset is a typical snapshot in the steady state.}
\label{fig:result2}
\end{figure}

\emph{Dynamic origin of MIPS}.--Firstly, we focus on the dynamics of the active system just before the MIPS threshold. As an example, the nonequilibrium potential of the active system with $v_0=40$ in the $\rho^{\prime}-E^{\prime}$ phase space is presented in Fig.~\ref{fig:result2}. It can be observed that there is only one potential well representing a single phase without any phase separation behaviors, which agrees with the typical snapshot shown in the inset of Fig.~\ref{fig:result2}. Moreover, the obtained flux field is further presented as the black arrows in Fig.~\ref{fig:result2}, where the nonzero flux demonstrates that the DB at the coarse-grained scale is broken. More interestingly, it can be found that the flux located inside the potential well can be explicitly divided into two parts with opposite directions. For a given $E^\prime$, the flux located in the region with small $\rho^{\prime}$ (on the left side of the purple dashed line in Fig.~\ref{fig:result2}) tends to push the local state to the negative direction of $\rho^{\prime}$, rendering $\rho^{\prime}$ to be increasingly lower. The flux located in the region with large $\rho^{\prime}$ (on the right side of the purple dashed line) prefers to push the local state to the positive direction of $\rho^{\prime}$, leading to increasingly higher $\rho^{\prime}$. From another view of a given $\rho^\prime$, the flux points to the lower $\rho^\prime$ direction when $E^\prime$ is large (above the purple dashed line) while the flux points to the higher $\rho^{\prime}$ direction when $E^\prime$ is small (below the purple dashed line). This means that the local state with high density or small average single-particle energy will be pushed to the higher $\rho^{\prime}$ direction by the nonequilibrium flux field, which is consistent with the self-trapping mechanism of MIPS proposed by Cates et al \cite{MIPS1}. In short, the flux field consisting of two parts with opposite directions tries to ``tear up" the potential well to create new wells with different densities. As mentioned, the driving force for the nonequilibrium dynamics can be decomposed into both gradient of the landscape and the curl flux. While the gradient force tends to attract the system down to the point attractor and stabilize it, the curl flux which breaks the DB is rotational. Thus, the rotational force tends to create flows rather than localizing at a point state. Therefore, the dynamical effect of the flux is to destabilize the point attractor state while stabilizing the continuous flow in the state space. Therefore, we hold the view that the nonequilibrium flux force is the dynamic origin of MIPS.

\begin{figure}
\centering
\includegraphics[width=1.0\columnwidth]{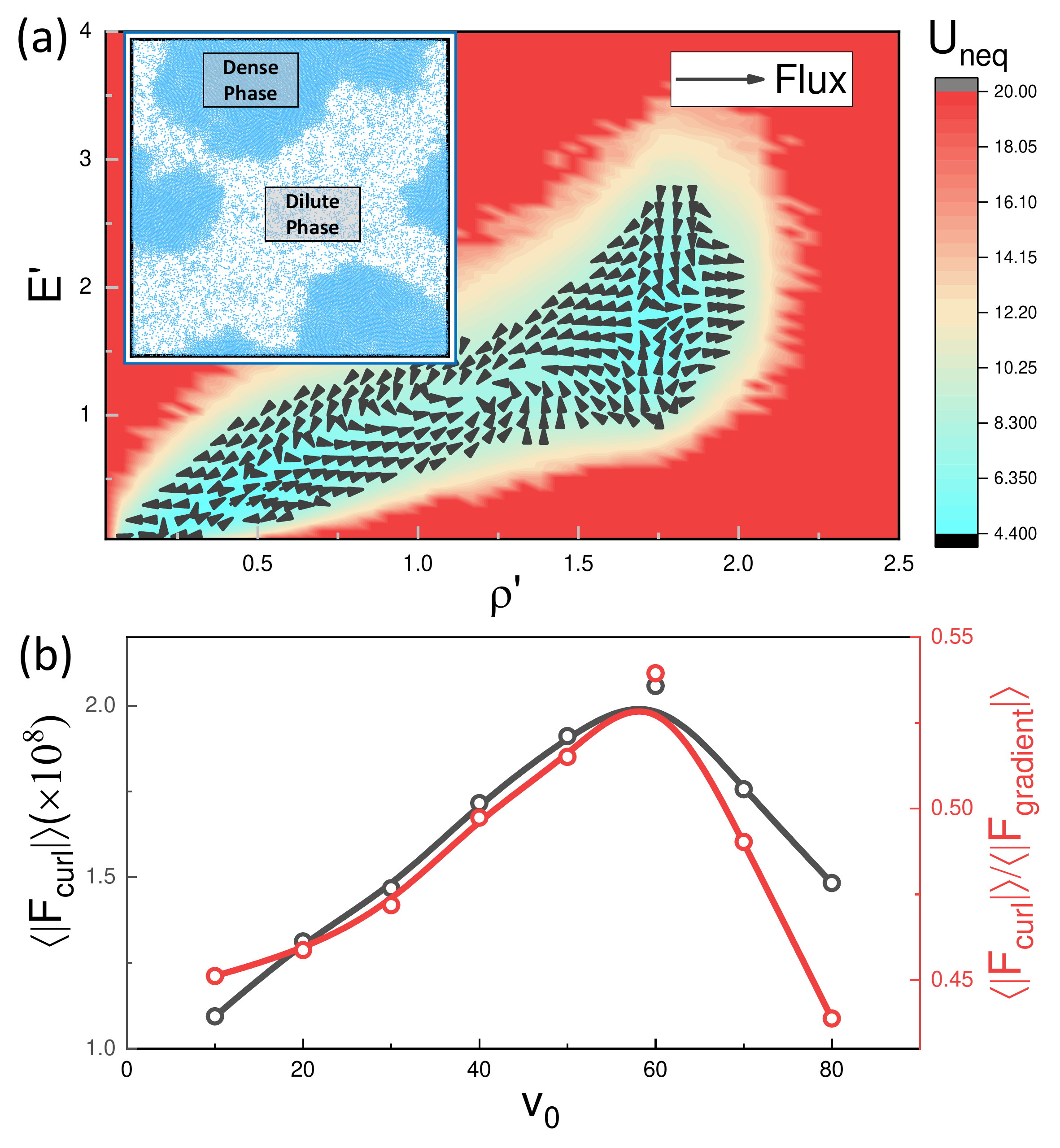}
\caption{(a) The nonequilibrium potential (colored background) and flux (black arrows) of the active system with $v_0=100$. The inset is a typical snapshot in the steady state. (b) Depedence of $\langle|\mathbf{F}_{curl}|\rangle$ and $\langle|\mathbf{F}_{curl}|\rangle/\langle|\mathbf{F}_{gradient}|\rangle$ on $v_0$.}
\label{fig:result3}
\end{figure}

Next, the dynamics after MIPS occurs (such as $v_0=100$) is considered. As presented in Fig.~\ref{fig:result3}(a), it can be observed that the nonequilibrium potential has two potential wells with different $\rho^{\prime}$, where the potential well located at small $\rho^{\prime}$ represents the dilute phase, while the one located at large $\rho^{\prime}$ denotes the dense phase, consistent with the typical snapshot shown in the inset of Fig.~\ref{fig:result3}(a). In addition, it is found that the flux rotates in a counterclockwise direction between these two potential wells, demonstrating again the DB breaking in the coarse-grained space due to nonequilibrium activity. It is noted that for the equilibrium phase separation without any nonequilibrium flux, particles in one of the potential wells are very difficult to jump into the other one with the sole help of noise. However, particles can be easily pushed from one phase to the other due to the nonequilibrium flux in active systems, which is consistent with the observations in simulations.

To quantitatively characterize the ``tearing effect" induced by the nonequilibrium flux in the MIPS process, we calculate the absolute value of the nonequilibrium flux and its relative value to the driving force due to nonequilibrium potential as the functions of activity $v_0$. That is, the curl part $\langle|\mathbf{F}_{curl}|\rangle$ as well as the ratio of the curl part to the gradient one $\langle|\mathbf{F}_{curl}|\rangle/\langle|\mathbf{F}_{gradient}|\rangle$, which can be calculated by $\langle|\mathbf{F}_{curl}|\rangle=\iint|\mathbf{J}_{ss}/P_{ss}|d\rho dE$, and $\langle|\mathbf{F}_{gradient}|\rangle=\iint|\mathbf{D}\cdot\nabla U_{neq}|d\rho dE$. The integral bounds of $\rho$ is set to be $[0.6,1.15]$ because only the flux located inside the regions with the density between the dense and dilute phases contributes to the formation of MIPS. The dependence of $\langle|\mathbf{F}_{curl}|\rangle$ and $\langle|\mathbf{F}_{curl}|\rangle/\langle|\mathbf{F}_{gradient}|\rangle$ on $v_0$ is illustrated in Fig.~\ref{fig:result3}(b). It can be found that both $\langle|\mathbf{F}_{curl}|\rangle$ (black symbols and line) and $\langle|\mathbf{F}_{curl}|\rangle/\langle|\mathbf{F}_{gradient}|\rangle$ (red symbols and line) depend nonmonotonically on $v_0$ and show maximal values around the MIPS threshold, further demonstrating that the nonequilibrium flux is the dynamic origin of the occurrence of MIPS.

\emph{Thermodynamic origin of MIPS}.--Now we focus on the thermodynamics of MIPS. To quantify DB breaking and time-reversal symmetry (TRS) in the density-energy phase space, we introduce the noise-averaged global entropy production rate $e_{p}$ in the steady states \cite{seifert2012stochastic}

\begin{equation}
\begin{aligned}
e_{p} &=\iint\frac{\mathbf{J}_{ss}^{{\rm T}}(\rho,E)\cdot\mathbf{D}^{-1}(\rho,E)\cdot\mathbf{J}_{ss}(\rho,E)}{P_{ss}(\rho,E)}d\rho dE \\
&=\iint\mathbf{v}_{ss}^{{\rm T}}(\rho,E)\cdot\mathbf{D}^{-1}(\rho,E)\cdot\mathbf{v}_{ss}(\rho,E)P_{ss}(\rho,E)d\rho dE.
\label{eq:EPR}
\end{aligned}
\end{equation}

\begin{figure}
\centering
\includegraphics[width=1.0\columnwidth]{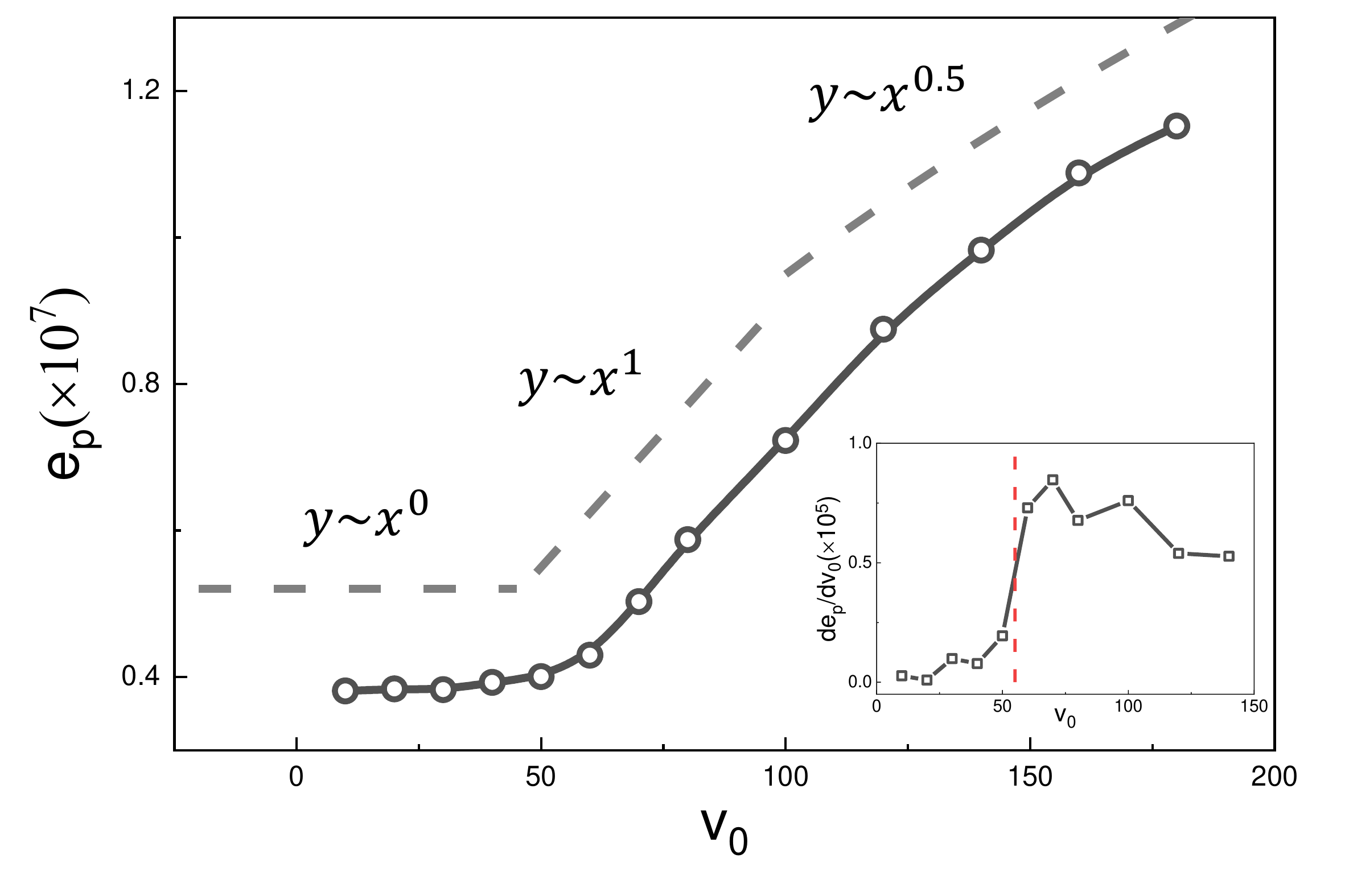}
\caption{Depedence of the entropy production rate $e_p$ on $v_0$. The dashed lines denote functions of scales 0, 1 and 0.5. The inset is the dependence of the derivative of $e_p$ on $v_0$ with the dashed red line the MIPS threshold.}
\label{fig:result5}
\end{figure}

The obtained $e_p$ dependent on the particle activity $v_0$ is shown in Fig.~\ref{fig:result5}. It can be found that $e_p$ increases monotonically as $v_0$ increases, indicating that higher activity will bring more dissipation into the system, further making the system far away from the equilibrium. More interestingly, it can be observed that $e_p$ scales differently on $v_0$ before and after the MIPS threshold. When $v_0$ is lower than the threshold, $e_p$ remains nearly unchanged. However, once $v_0$ exceeds the threshold $e_p$ increases linearly with a slope of approximately 1. To take a closer look at this transition, the obtained derivative of $e_p$ dependent on $v_0$ is illustrated in the inset of Fig.~\ref{fig:result5}. Clearly, $de_p/dv_0$ holds slightly higher than $0$ before the MIPS threshold (the dashed red line), and rapidly becomes very large when $v_0$ passes the threshold. Since the steady state flux is mainly contributed by the $E$ component of the local flow (see SI for details), the net transition in the energy space dominates the scaling behavior and sharpness of entropy production. The thermodynamic origin of MIPS can be attributed to the sharp transition of scaling behavior of the entropy production in the density-energy phase space. From the expression of the entropy production, it is directly related to the flux measuring the degree of DB breaking. The flux thus forms the dynamical basis of the nonequilibrium thermodynamic cost. Since there is a steep change of nonequilibrium thermodynamic cost characterized by EPR when the MIPS system transforms from a single phase (one attractor basin) to phase separation with two coexisting phases (two attractor basins), showing that the EPR may provide the thermodynamic origin of MIPS. In addition, this observation provides a reliable tool to identify the MIPS threshold between the single and coexisting phases, which confirms that the phase transition point can be characterized in terms of entropy production \cite{TP1,TP2,TP3,TP4,TP5}, especially in active matter systems \cite{TPA1,TPA2,TPA3}. Compared with the other studies focusing on the entropy production of active matter \cite{ep1,ep2,ep3,ep4,ep5,ep6,ep7,ep8,ep9,ep10,ep11,ep12,ep13,ep14,ep15}, our findings provide fresh insights into the origin of MIPS.

In summary, the dynamic and thermodynamic origins of MIPS are investigated by the nonequilibrium potential and flux landscape theory in the density-energy phase space through the brand-new coarse-grained mapping method we proposed. As a result, it is found that the nonequilibrium flux with opposite directions tends to ``tear up" the single nonequilibrium potential well before the MIPS threshold. This not only demonstrates directly that the nonequilibrium flux is the dynamic origin of the occurrence of MIPS but also provides evidence that the DB is also significantly broken at the coarse-grained scale. In addition, intensive simulations reveal that the EPR shows a transition of scaling behavior around the MIPS threshold, which serves as the thermodynamic origin of MIPS. Our findings bring new insights and propose a new route to understand the nonequilibrium nature of MIPS.

J. Su, Z. Cao, H. Jiang and Z. Hou are supported by MOST(2018YFA0208702), NSFC (32090044, 21973085, 21833007), Anhui Initiative in Quantum Information Technologies (AHY090200), and the Fundamental Research Funds for the Central Universities (WK2340000104).


\begin{thebibliography}{79}
\expandafter\ifx\csname natexlab\endcsname\relax\def\natexlab#1{#1}\fi
\expandafter\ifx\csname bibnamefont\endcsname\relax
  \def\bibnamefont#1{#1}\fi
\expandafter\ifx\csname bibfnamefont\endcsname\relax
  \def\bibfnamefont#1{#1}\fi
\expandafter\ifx\csname citenamefont\endcsname\relax
  \def\citenamefont#1{#1}\fi
\expandafter\ifx\csname url\endcsname\relax
  \def\url#1{\texttt{#1}}\fi
\expandafter\ifx\csname urlprefix\endcsname\relax\def\urlprefix{URL }\fi
\providecommand{\bibinfo}[2]{#2}
\providecommand{\eprint}[2][]{\url{#2}}

\bibitem[{\citenamefont{Bechinger et~al.}(2016)\citenamefont{Bechinger,
  Di~Leonardo, L{\"o}wen, Reichhardt, Volpe, and Volpe}}]{a1}
\bibinfo{author}{\bibfnamefont{C.}~\bibnamefont{Bechinger}},
  \bibinfo{author}{\bibfnamefont{R.}~\bibnamefont{Di~Leonardo}},
  \bibinfo{author}{\bibfnamefont{H.}~\bibnamefont{L{\"o}wen}},
  \bibinfo{author}{\bibfnamefont{C.}~\bibnamefont{Reichhardt}},
  \bibinfo{author}{\bibfnamefont{G.}~\bibnamefont{Volpe}}, \bibnamefont{and}
  \bibinfo{author}{\bibfnamefont{G.}~\bibnamefont{Volpe}},
  \bibinfo{journal}{Rev. Mod. Phys.} \textbf{\bibinfo{volume}{88}},
  \bibinfo{pages}{045006} (\bibinfo{year}{2016}).

\bibitem[{\citenamefont{Vicsek and Zafeiris}(2012)}]{a2}
\bibinfo{author}{\bibfnamefont{T.}~\bibnamefont{Vicsek}} \bibnamefont{and}
  \bibinfo{author}{\bibfnamefont{A.}~\bibnamefont{Zafeiris}},
  \bibinfo{journal}{Phys. Rep.} \textbf{\bibinfo{volume}{517}},
  \bibinfo{pages}{71} (\bibinfo{year}{2012}), ISSN \bibinfo{issn}{0370-1573}.

\bibitem[{\citenamefont{Battle et~al.}(2016)\citenamefont{Battle, Broedersz,
  Fakhri, Geyer, Howard, Schmidt, and MacKintosh}}]{B1}
\bibinfo{author}{\bibfnamefont{C.}~\bibnamefont{Battle}},
  \bibinfo{author}{\bibfnamefont{C.~P.} \bibnamefont{Broedersz}},
  \bibinfo{author}{\bibfnamefont{N.}~\bibnamefont{Fakhri}},
  \bibinfo{author}{\bibfnamefont{V.~F.} \bibnamefont{Geyer}},
  \bibinfo{author}{\bibfnamefont{J.}~\bibnamefont{Howard}},
  \bibinfo{author}{\bibfnamefont{C.~F.} \bibnamefont{Schmidt}},
  \bibnamefont{and} \bibinfo{author}{\bibfnamefont{F.~C.}
  \bibnamefont{MacKintosh}}, \bibinfo{journal}{Science}
  \textbf{\bibinfo{volume}{352}}, \bibinfo{pages}{604} (\bibinfo{year}{2016}).

\bibitem[{\citenamefont{Gnesotto et~al.}(2018)\citenamefont{Gnesotto, Mura,
  Gladrow, and Broedersz}}]{B2}
\bibinfo{author}{\bibfnamefont{F.~S.} \bibnamefont{Gnesotto}},
  \bibinfo{author}{\bibfnamefont{F.}~\bibnamefont{Mura}},
  \bibinfo{author}{\bibfnamefont{J.}~\bibnamefont{Gladrow}}, \bibnamefont{and}
  \bibinfo{author}{\bibfnamefont{C.~P.} \bibnamefont{Broedersz}},
  \bibinfo{journal}{Rep. Prog. Phys.} \textbf{\bibinfo{volume}{81}},
  \bibinfo{pages}{066601} (\bibinfo{year}{2018}).

\bibitem[{\citenamefont{Gladrow et~al.}(2016)\citenamefont{Gladrow, Fakhri,
  MacKintosh, Schmidt, and Broedersz}}]{B3}
\bibinfo{author}{\bibfnamefont{J.}~\bibnamefont{Gladrow}},
  \bibinfo{author}{\bibfnamefont{N.}~\bibnamefont{Fakhri}},
  \bibinfo{author}{\bibfnamefont{F.~C.} \bibnamefont{MacKintosh}},
  \bibinfo{author}{\bibfnamefont{C.}~\bibnamefont{Schmidt}}, \bibnamefont{and}
  \bibinfo{author}{\bibfnamefont{C.}~\bibnamefont{Broedersz}},
  \bibinfo{journal}{Phys. Rev. Lett.} \textbf{\bibinfo{volume}{116}},
  \bibinfo{pages}{248301} (\bibinfo{year}{2016}).

\bibitem[{\citenamefont{Mura et~al.}(2018)\citenamefont{Mura, Gradziuk, and
  Broedersz}}]{B4}
\bibinfo{author}{\bibfnamefont{F.}~\bibnamefont{Mura}},
  \bibinfo{author}{\bibfnamefont{G.}~\bibnamefont{Gradziuk}}, \bibnamefont{and}
  \bibinfo{author}{\bibfnamefont{C.~P.} \bibnamefont{Broedersz}},
  \bibinfo{journal}{Phys. Rev. Lett.} \textbf{\bibinfo{volume}{121}},
  \bibinfo{pages}{038002} (\bibinfo{year}{2018}).

\bibitem[{\citenamefont{Seara et~al.}(2018)\citenamefont{Seara, Yadav,
  Linsmeier, Tabatabai, Oakes, Tabei, Banerjee, and Murrell}}]{B5}
\bibinfo{author}{\bibfnamefont{D.~S.} \bibnamefont{Seara}},
  \bibinfo{author}{\bibfnamefont{V.}~\bibnamefont{Yadav}},
  \bibinfo{author}{\bibfnamefont{I.}~\bibnamefont{Linsmeier}},
  \bibinfo{author}{\bibfnamefont{A.~P.} \bibnamefont{Tabatabai}},
  \bibinfo{author}{\bibfnamefont{P.~W.} \bibnamefont{Oakes}},
  \bibinfo{author}{\bibfnamefont{S.}~\bibnamefont{Tabei}},
  \bibinfo{author}{\bibfnamefont{S.}~\bibnamefont{Banerjee}}, \bibnamefont{and}
  \bibinfo{author}{\bibfnamefont{M.~P.} \bibnamefont{Murrell}},
  \bibinfo{journal}{Nat. Commun.} \textbf{\bibinfo{volume}{9}},
  \bibinfo{pages}{1} (\bibinfo{year}{2018}).

\bibitem[{\citenamefont{Gladrow et~al.}(2017)\citenamefont{Gladrow, Broedersz,
  and Schmidt}}]{B6}
\bibinfo{author}{\bibfnamefont{J.}~\bibnamefont{Gladrow}},
  \bibinfo{author}{\bibfnamefont{C.~P.} \bibnamefont{Broedersz}},
  \bibnamefont{and} \bibinfo{author}{\bibfnamefont{C.~F.}
  \bibnamefont{Schmidt}}, \bibinfo{journal}{Phys. Rev. E}
  \textbf{\bibinfo{volume}{96}}, \bibinfo{pages}{022408}
  (\bibinfo{year}{2017}).

\bibitem[{\citenamefont{Dieball and Godec}(2022{\natexlab{a}})}]{B7}
\bibinfo{author}{\bibfnamefont{C.}~\bibnamefont{Dieball}} \bibnamefont{and}
  \bibinfo{author}{\bibfnamefont{A.}~\bibnamefont{Godec}},
  \bibinfo{journal}{Phys. Rev. Lett.} \textbf{\bibinfo{volume}{129}},
  \bibinfo{pages}{140601} (\bibinfo{year}{2022}{\natexlab{a}}).

\bibitem[{\citenamefont{Dieball and Godec}(2022{\natexlab{b}})}]{B8}
\bibinfo{author}{\bibfnamefont{C.}~\bibnamefont{Dieball}} \bibnamefont{and}
  \bibinfo{author}{\bibfnamefont{A.}~\bibnamefont{Godec}},
  \bibinfo{journal}{arXiv preprint arXiv:2204.06553}
  (\bibinfo{year}{2022}{\natexlab{b}}).

\bibitem[{\citenamefont{DiLuzio et~al.}(2005)\citenamefont{DiLuzio, Turner,
  Mayer, Garstecki, Weibel, Berg, and Whitesides}}]{a4}
\bibinfo{author}{\bibfnamefont{W.~R.} \bibnamefont{DiLuzio}},
  \bibinfo{author}{\bibfnamefont{L.}~\bibnamefont{Turner}},
  \bibinfo{author}{\bibfnamefont{M.}~\bibnamefont{Mayer}},
  \bibinfo{author}{\bibfnamefont{P.}~\bibnamefont{Garstecki}},
  \bibinfo{author}{\bibfnamefont{D.~B.} \bibnamefont{Weibel}},
  \bibinfo{author}{\bibfnamefont{H.~C.} \bibnamefont{Berg}}, \bibnamefont{and}
  \bibinfo{author}{\bibfnamefont{G.~M.} \bibnamefont{Whitesides}},
  \bibinfo{journal}{Nature} \textbf{\bibinfo{volume}{435}},
  \bibinfo{pages}{1271} (\bibinfo{year}{2005}).

\bibitem[{\citenamefont{Riedel et~al.}(2005)\citenamefont{Riedel, Kruse, and
  Howard}}]{a5}
\bibinfo{author}{\bibfnamefont{I.~H.} \bibnamefont{Riedel}},
  \bibinfo{author}{\bibfnamefont{K.}~\bibnamefont{Kruse}}, \bibnamefont{and}
  \bibinfo{author}{\bibfnamefont{J.}~\bibnamefont{Howard}},
  \bibinfo{journal}{Science} \textbf{\bibinfo{volume}{309}},
  \bibinfo{pages}{300} (\bibinfo{year}{2005}).

\bibitem[{\citenamefont{K{\"u}mmel et~al.}(2013)\citenamefont{K{\"u}mmel, ten
  Hagen, Wittkowski, Buttinoni, Eichhorn, Volpe, L{\"o}wen, and
  Bechinger}}]{a6}
\bibinfo{author}{\bibfnamefont{F.}~\bibnamefont{K{\"u}mmel}},
  \bibinfo{author}{\bibfnamefont{B.}~\bibnamefont{ten Hagen}},
  \bibinfo{author}{\bibfnamefont{R.}~\bibnamefont{Wittkowski}},
  \bibinfo{author}{\bibfnamefont{I.}~\bibnamefont{Buttinoni}},
  \bibinfo{author}{\bibfnamefont{R.}~\bibnamefont{Eichhorn}},
  \bibinfo{author}{\bibfnamefont{G.}~\bibnamefont{Volpe}},
  \bibinfo{author}{\bibfnamefont{H.}~\bibnamefont{L{\"o}wen}},
  \bibnamefont{and}
  \bibinfo{author}{\bibfnamefont{C.}~\bibnamefont{Bechinger}},
  \bibinfo{journal}{Phys. Rev. Lett.} \textbf{\bibinfo{volume}{110}},
  \bibinfo{pages}{198302} (\bibinfo{year}{2013}).

\bibitem[{\citenamefont{Gibbs et~al.}(2011)\citenamefont{Gibbs, Kothari,
  Saintillan, and Zhao}}]{a7}
\bibinfo{author}{\bibfnamefont{J.}~\bibnamefont{Gibbs}},
  \bibinfo{author}{\bibfnamefont{S.}~\bibnamefont{Kothari}},
  \bibinfo{author}{\bibfnamefont{D.}~\bibnamefont{Saintillan}},
  \bibnamefont{and} \bibinfo{author}{\bibfnamefont{Y.-P.} \bibnamefont{Zhao}},
  \bibinfo{journal}{Nano Lett.} \textbf{\bibinfo{volume}{11}},
  \bibinfo{pages}{2543} (\bibinfo{year}{2011}).

\bibitem[{\citenamefont{Mallory and Cacciuto}(2019)}]{a8}
\bibinfo{author}{\bibfnamefont{S.~A.} \bibnamefont{Mallory}} \bibnamefont{and}
  \bibinfo{author}{\bibfnamefont{A.}~\bibnamefont{Cacciuto}},
  \bibinfo{journal}{J. Am. Chem. Soc.} \textbf{\bibinfo{volume}{141}},
  \bibinfo{pages}{2500} (\bibinfo{year}{2019}).

\bibitem[{\citenamefont{Gou et~al.}(2019)\citenamefont{Gou, Jiang, and
  Hou}}]{a9}
\bibinfo{author}{\bibfnamefont{Y.}~\bibnamefont{Gou}},
  \bibinfo{author}{\bibfnamefont{H.}~\bibnamefont{Jiang}}, \bibnamefont{and}
  \bibinfo{author}{\bibfnamefont{Z.}~\bibnamefont{Hou}}, \bibinfo{journal}{Soft
  Matter} \textbf{\bibinfo{volume}{15}}, \bibinfo{pages}{9104}
  (\bibinfo{year}{2019}).

\bibitem[{\citenamefont{Du et~al.}(2019)\citenamefont{Du, Jiang, and
  Hou}}]{a10}
\bibinfo{author}{\bibfnamefont{Y.}~\bibnamefont{Du}},
  \bibinfo{author}{\bibfnamefont{H.}~\bibnamefont{Jiang}}, \bibnamefont{and}
  \bibinfo{author}{\bibfnamefont{Z.}~\bibnamefont{Hou}}, \bibinfo{journal}{J.
  Chem. Phys.} \textbf{\bibinfo{volume}{151}}, \bibinfo{pages}{154904}
  (\bibinfo{year}{2019}).

\bibitem[{\citenamefont{Yan et~al.}(2016)\citenamefont{Yan, Han, Zhang, Xu,
  Luijten, and Granick}}]{a11}
\bibinfo{author}{\bibfnamefont{J.}~\bibnamefont{Yan}},
  \bibinfo{author}{\bibfnamefont{M.}~\bibnamefont{Han}},
  \bibinfo{author}{\bibfnamefont{J.}~\bibnamefont{Zhang}},
  \bibinfo{author}{\bibfnamefont{C.}~\bibnamefont{Xu}},
  \bibinfo{author}{\bibfnamefont{E.}~\bibnamefont{Luijten}}, \bibnamefont{and}
  \bibinfo{author}{\bibfnamefont{S.}~\bibnamefont{Granick}},
  \bibinfo{journal}{Nat. Mater.} \textbf{\bibinfo{volume}{15}},
  \bibinfo{pages}{1095} (\bibinfo{year}{2016}).

\bibitem[{\citenamefont{Sumino et~al.}(2012)\citenamefont{Sumino, Nagai,
  Shitaka, Tanaka, Yoshikawa, Chat{\'e}, and Oiwa}}]{a12}
\bibinfo{author}{\bibfnamefont{Y.}~\bibnamefont{Sumino}},
  \bibinfo{author}{\bibfnamefont{K.~H.} \bibnamefont{Nagai}},
  \bibinfo{author}{\bibfnamefont{Y.}~\bibnamefont{Shitaka}},
  \bibinfo{author}{\bibfnamefont{D.}~\bibnamefont{Tanaka}},
  \bibinfo{author}{\bibfnamefont{K.}~\bibnamefont{Yoshikawa}},
  \bibinfo{author}{\bibfnamefont{H.}~\bibnamefont{Chat{\'e}}},
  \bibnamefont{and} \bibinfo{author}{\bibfnamefont{K.}~\bibnamefont{Oiwa}},
  \bibinfo{journal}{Nature} \textbf{\bibinfo{volume}{483}},
  \bibinfo{pages}{448} (\bibinfo{year}{2012}).

\bibitem[{\citenamefont{Jiang et~al.}(2017)\citenamefont{Jiang, Ding, Pu, and
  Hou}}]{a13}
\bibinfo{author}{\bibfnamefont{H.}~\bibnamefont{Jiang}},
  \bibinfo{author}{\bibfnamefont{H.}~\bibnamefont{Ding}},
  \bibinfo{author}{\bibfnamefont{M.}~\bibnamefont{Pu}}, \bibnamefont{and}
  \bibinfo{author}{\bibfnamefont{Z.}~\bibnamefont{Hou}}, \bibinfo{journal}{Soft
  matter} \textbf{\bibinfo{volume}{13}}, \bibinfo{pages}{836}
  (\bibinfo{year}{2017}).

\bibitem[{\citenamefont{Karani et~al.}(2019)\citenamefont{Karani, Pradillo, and
  Vlahovska}}]{a14}
\bibinfo{author}{\bibfnamefont{H.}~\bibnamefont{Karani}},
  \bibinfo{author}{\bibfnamefont{G.~E.} \bibnamefont{Pradillo}},
  \bibnamefont{and} \bibinfo{author}{\bibfnamefont{P.~M.}
  \bibnamefont{Vlahovska}}, \bibinfo{journal}{Phys. Rev. Lett.}
  \textbf{\bibinfo{volume}{123}}, \bibinfo{pages}{208002}
  (\bibinfo{year}{2019}).

\bibitem[{\citenamefont{Gou et~al.}(2020)\citenamefont{Gou, Jiang, and
  Hou}}]{a15}
\bibinfo{author}{\bibfnamefont{Y.-l.} \bibnamefont{Gou}},
  \bibinfo{author}{\bibfnamefont{H.-j.} \bibnamefont{Jiang}}, \bibnamefont{and}
  \bibinfo{author}{\bibfnamefont{Z.-h.} \bibnamefont{Hou}},
  \bibinfo{journal}{Chin. J. Chem. Phys.} \textbf{\bibinfo{volume}{33}},
  \bibinfo{pages}{717} (\bibinfo{year}{2020}).

\bibitem[{\citenamefont{Tailleur and Cates}(2008)}]{MIPS1}
\bibinfo{author}{\bibfnamefont{J.}~\bibnamefont{Tailleur}} \bibnamefont{and}
  \bibinfo{author}{\bibfnamefont{M.~E.} \bibnamefont{Cates}},
  \bibinfo{journal}{Phys. Rev. Lett.} \textbf{\bibinfo{volume}{100}},
  \bibinfo{pages}{218103} (\bibinfo{year}{2008}).

\bibitem[{\citenamefont{Cates and Tailleur}(2015)}]{MIPS2}
\bibinfo{author}{\bibfnamefont{M.~E.} \bibnamefont{Cates}} \bibnamefont{and}
  \bibinfo{author}{\bibfnamefont{J.}~\bibnamefont{Tailleur}},
  \bibinfo{journal}{Annu. Rev. Condens. Matter Phys.}
  \textbf{\bibinfo{volume}{6}}, \bibinfo{pages}{219} (\bibinfo{year}{2015}).

\bibitem[{\citenamefont{Redner et~al.}(2013{\natexlab{a}})\citenamefont{Redner,
  Hagan, and Baskaran}}]{MIPS3}
\bibinfo{author}{\bibfnamefont{G.~S.} \bibnamefont{Redner}},
  \bibinfo{author}{\bibfnamefont{M.~F.} \bibnamefont{Hagan}}, \bibnamefont{and}
  \bibinfo{author}{\bibfnamefont{A.}~\bibnamefont{Baskaran}},
  \bibinfo{journal}{Phys. Rev. Lett.} \textbf{\bibinfo{volume}{110}},
  \bibinfo{pages}{055701} (\bibinfo{year}{2013}{\natexlab{a}}).

\bibitem[{\citenamefont{Redner et~al.}(2013{\natexlab{b}})\citenamefont{Redner,
  Baskaran, and Hagan}}]{MIPS4}
\bibinfo{author}{\bibfnamefont{G.~S.} \bibnamefont{Redner}},
  \bibinfo{author}{\bibfnamefont{A.}~\bibnamefont{Baskaran}}, \bibnamefont{and}
  \bibinfo{author}{\bibfnamefont{M.~F.} \bibnamefont{Hagan}},
  \bibinfo{journal}{Phys. Rev. E} \textbf{\bibinfo{volume}{88}},
  \bibinfo{pages}{012305} (\bibinfo{year}{2013}{\natexlab{b}}).

\bibitem[{\citenamefont{Redner et~al.}(2016)\citenamefont{Redner, Wagner,
  Baskaran, and Hagan}}]{MIPS5}
\bibinfo{author}{\bibfnamefont{G.~S.} \bibnamefont{Redner}},
  \bibinfo{author}{\bibfnamefont{C.~G.} \bibnamefont{Wagner}},
  \bibinfo{author}{\bibfnamefont{A.}~\bibnamefont{Baskaran}}, \bibnamefont{and}
  \bibinfo{author}{\bibfnamefont{M.~F.} \bibnamefont{Hagan}},
  \bibinfo{journal}{Phys. Rev. Lett.} \textbf{\bibinfo{volume}{117}},
  \bibinfo{pages}{148002} (\bibinfo{year}{2016}).

\bibitem[{\citenamefont{Takatori et~al.}(2014)\citenamefont{Takatori, Yan, and
  Brady}}]{MIPS11}
\bibinfo{author}{\bibfnamefont{S.~C.} \bibnamefont{Takatori}},
  \bibinfo{author}{\bibfnamefont{W.}~\bibnamefont{Yan}}, \bibnamefont{and}
  \bibinfo{author}{\bibfnamefont{J.~F.} \bibnamefont{Brady}},
  \bibinfo{journal}{Phys. Rev. Lett.} \textbf{\bibinfo{volume}{113}},
  \bibinfo{pages}{028103} (\bibinfo{year}{2014}).

\bibitem[{\citenamefont{Takatori and Brady}(2015)}]{MIPS12}
\bibinfo{author}{\bibfnamefont{S.~C.} \bibnamefont{Takatori}} \bibnamefont{and}
  \bibinfo{author}{\bibfnamefont{J.~F.} \bibnamefont{Brady}},
  \bibinfo{journal}{Phys. Rev. E} \textbf{\bibinfo{volume}{91}},
  \bibinfo{pages}{032117} (\bibinfo{year}{2015}).

\bibitem[{\citenamefont{Patch et~al.}(2017)\citenamefont{Patch, Yllanes, and
  Marchetti}}]{MIPS13}
\bibinfo{author}{\bibfnamefont{A.}~\bibnamefont{Patch}},
  \bibinfo{author}{\bibfnamefont{D.}~\bibnamefont{Yllanes}}, \bibnamefont{and}
  \bibinfo{author}{\bibfnamefont{M.~C.} \bibnamefont{Marchetti}},
  \bibinfo{journal}{Phys. Rev. E} \textbf{\bibinfo{volume}{95}},
  \bibinfo{pages}{012601} (\bibinfo{year}{2017}).

\bibitem[{\citenamefont{Fily et~al.}(2014)\citenamefont{Fily, Henkes, and
  Marchetti}}]{MIPS14}
\bibinfo{author}{\bibfnamefont{Y.}~\bibnamefont{Fily}},
  \bibinfo{author}{\bibfnamefont{S.}~\bibnamefont{Henkes}}, \bibnamefont{and}
  \bibinfo{author}{\bibfnamefont{M.~C.} \bibnamefont{Marchetti}},
  \bibinfo{journal}{Soft Matter} \textbf{\bibinfo{volume}{10}},
  \bibinfo{pages}{2132} (\bibinfo{year}{2014}).

\bibitem[{\citenamefont{Stenhammar et~al.}(2014)\citenamefont{Stenhammar,
  Marenduzzo, Allen, and Cates}}]{MIPS15}
\bibinfo{author}{\bibfnamefont{J.}~\bibnamefont{Stenhammar}},
  \bibinfo{author}{\bibfnamefont{D.}~\bibnamefont{Marenduzzo}},
  \bibinfo{author}{\bibfnamefont{R.~J.} \bibnamefont{Allen}}, \bibnamefont{and}
  \bibinfo{author}{\bibfnamefont{M.~E.} \bibnamefont{Cates}},
  \bibinfo{journal}{Soft Matter} \textbf{\bibinfo{volume}{10}},
  \bibinfo{pages}{1489} (\bibinfo{year}{2014}).

\bibitem[{\citenamefont{Z\"ottl and Stark}(2014)}]{MIPS16}
\bibinfo{author}{\bibfnamefont{A.}~\bibnamefont{Z\"ottl}} \bibnamefont{and}
  \bibinfo{author}{\bibfnamefont{H.}~\bibnamefont{Stark}},
  \bibinfo{journal}{Phys. Rev. Lett.} \textbf{\bibinfo{volume}{112}},
  \bibinfo{pages}{118101} (\bibinfo{year}{2014}).

\bibitem[{\citenamefont{Furukawa et~al.}(2014)\citenamefont{Furukawa,
  Marenduzzo, and Cates}}]{MIPS17}
\bibinfo{author}{\bibfnamefont{A.}~\bibnamefont{Furukawa}},
  \bibinfo{author}{\bibfnamefont{D.}~\bibnamefont{Marenduzzo}},
  \bibnamefont{and} \bibinfo{author}{\bibfnamefont{M.~E.} \bibnamefont{Cates}},
  \bibinfo{journal}{Phys. Rev. E} \textbf{\bibinfo{volume}{90}},
  \bibinfo{pages}{022303} (\bibinfo{year}{2014}).

\bibitem[{\citenamefont{Blaschke et~al.}(2016)\citenamefont{Blaschke, Maurer,
  Menon, Zöttl, and Stark}}]{MIPS18}
\bibinfo{author}{\bibfnamefont{J.}~\bibnamefont{Blaschke}},
  \bibinfo{author}{\bibfnamefont{M.}~\bibnamefont{Maurer}},
  \bibinfo{author}{\bibfnamefont{K.}~\bibnamefont{Menon}},
  \bibinfo{author}{\bibfnamefont{A.}~\bibnamefont{Zöttl}}, \bibnamefont{and}
  \bibinfo{author}{\bibfnamefont{H.}~\bibnamefont{Stark}},
  \bibinfo{journal}{Soft Matter} \textbf{\bibinfo{volume}{12}},
  \bibinfo{pages}{9821} (\bibinfo{year}{2016}).

\bibitem[{\citenamefont{Stenhammar et~al.}(2015)\citenamefont{Stenhammar,
  Wittkowski, Marenduzzo, and Cates}}]{MIPS19}
\bibinfo{author}{\bibfnamefont{J.}~\bibnamefont{Stenhammar}},
  \bibinfo{author}{\bibfnamefont{R.}~\bibnamefont{Wittkowski}},
  \bibinfo{author}{\bibfnamefont{D.}~\bibnamefont{Marenduzzo}},
  \bibnamefont{and} \bibinfo{author}{\bibfnamefont{M.~E.} \bibnamefont{Cates}},
  \bibinfo{journal}{Phys. Rev. Lett.} \textbf{\bibinfo{volume}{114}},
  \bibinfo{pages}{018301} (\bibinfo{year}{2015}).

\bibitem[{\citenamefont{Mandal et~al.}(2019)\citenamefont{Mandal, Liebchen, and
  L\"owen}}]{MIPS20}
\bibinfo{author}{\bibfnamefont{S.}~\bibnamefont{Mandal}},
  \bibinfo{author}{\bibfnamefont{B.}~\bibnamefont{Liebchen}}, \bibnamefont{and}
  \bibinfo{author}{\bibfnamefont{H.}~\bibnamefont{L\"owen}},
  \bibinfo{journal}{Phys. Rev. Lett.} \textbf{\bibinfo{volume}{123}},
  \bibinfo{pages}{228001} (\bibinfo{year}{2019}).

\bibitem[{\citenamefont{Siebert et~al.}(2017)\citenamefont{Siebert, Letz,
  Speck, and Virnau}}]{MIPS21}
\bibinfo{author}{\bibfnamefont{J.~T.} \bibnamefont{Siebert}},
  \bibinfo{author}{\bibfnamefont{J.}~\bibnamefont{Letz}},
  \bibinfo{author}{\bibfnamefont{T.}~\bibnamefont{Speck}}, \bibnamefont{and}
  \bibinfo{author}{\bibfnamefont{P.}~\bibnamefont{Virnau}},
  \bibinfo{journal}{Soft Matter} \textbf{\bibinfo{volume}{13}},
  \bibinfo{pages}{1020} (\bibinfo{year}{2017}).

\bibitem[{\citenamefont{Liao and Klapp}(2018)}]{MIPS22}
\bibinfo{author}{\bibfnamefont{G.-J.} \bibnamefont{Liao}} \bibnamefont{and}
  \bibinfo{author}{\bibfnamefont{S.~H.~L.} \bibnamefont{Klapp}},
  \bibinfo{journal}{Soft Matter} \textbf{\bibinfo{volume}{14}},
  \bibinfo{pages}{7873} (\bibinfo{year}{2018}).

\bibitem[{\citenamefont{Su et~al.}(2021)\citenamefont{Su, Jiang, and
  Hou}}]{MIPS23}
\bibinfo{author}{\bibfnamefont{J.}~\bibnamefont{Su}},
  \bibinfo{author}{\bibfnamefont{H.}~\bibnamefont{Jiang}}, \bibnamefont{and}
  \bibinfo{author}{\bibfnamefont{Z.}~\bibnamefont{Hou}}, \bibinfo{journal}{New
  J. Phys.} \textbf{\bibinfo{volume}{23}}, \bibinfo{pages}{013005}
  (\bibinfo{year}{2021}).

\bibitem[{\citenamefont{Du et~al.}(2020)\citenamefont{Du, Jiang, and
  Hou}}]{MIPS24}
\bibinfo{author}{\bibfnamefont{Y.}~\bibnamefont{Du}},
  \bibinfo{author}{\bibfnamefont{H.}~\bibnamefont{Jiang}}, \bibnamefont{and}
  \bibinfo{author}{\bibfnamefont{Z.}~\bibnamefont{Hou}}, \bibinfo{journal}{Soft
  Matter} \textbf{\bibinfo{volume}{16}}, \bibinfo{pages}{6434}
  (\bibinfo{year}{2020}).

\bibitem[{\citenamefont{Caprini
  et~al.}(2020{\natexlab{a}})\citenamefont{Caprini, Marini Bettolo~Marconi, and
  Puglisi}}]{MIPS25}
\bibinfo{author}{\bibfnamefont{L.}~\bibnamefont{Caprini}},
  \bibinfo{author}{\bibfnamefont{U.}~\bibnamefont{Marini Bettolo~Marconi}},
  \bibnamefont{and} \bibinfo{author}{\bibfnamefont{A.}~\bibnamefont{Puglisi}},
  \bibinfo{journal}{Phys. Rev. Lett.} \textbf{\bibinfo{volume}{124}},
  \bibinfo{pages}{078001} (\bibinfo{year}{2020}{\natexlab{a}}).

\bibitem[{\citenamefont{Caprini
  et~al.}(2020{\natexlab{b}})\citenamefont{Caprini, Marini Bettolo~Marconi,
  Maggi, Paoluzzi, and Puglisi}}]{MIPS26}
\bibinfo{author}{\bibfnamefont{L.}~\bibnamefont{Caprini}},
  \bibinfo{author}{\bibfnamefont{U.}~\bibnamefont{Marini Bettolo~Marconi}},
  \bibinfo{author}{\bibfnamefont{C.}~\bibnamefont{Maggi}},
  \bibinfo{author}{\bibfnamefont{M.}~\bibnamefont{Paoluzzi}}, \bibnamefont{and}
  \bibinfo{author}{\bibfnamefont{A.}~\bibnamefont{Puglisi}},
  \bibinfo{journal}{Phys. Rev. Research} \textbf{\bibinfo{volume}{2}},
  \bibinfo{pages}{023321} (\bibinfo{year}{2020}{\natexlab{b}}).

\bibitem[{\citenamefont{Wittkowski et~al.}(2014)\citenamefont{Wittkowski,
  Tiribocchi, Stenhammar, Allen, Marenduzzo, and Cates}}]{MIPS9}
\bibinfo{author}{\bibfnamefont{R.}~\bibnamefont{Wittkowski}},
  \bibinfo{author}{\bibfnamefont{A.}~\bibnamefont{Tiribocchi}},
  \bibinfo{author}{\bibfnamefont{J.}~\bibnamefont{Stenhammar}},
  \bibinfo{author}{\bibfnamefont{R.~J.} \bibnamefont{Allen}},
  \bibinfo{author}{\bibfnamefont{D.}~\bibnamefont{Marenduzzo}},
  \bibnamefont{and} \bibinfo{author}{\bibfnamefont{M.~E.} \bibnamefont{Cates}},
  \bibinfo{journal}{Nat. Commun.} \textbf{\bibinfo{volume}{5}},
  \bibinfo{pages}{4351} (\bibinfo{year}{2014}).

\bibitem[{\citenamefont{Tjhung et~al.}(2018)\citenamefont{Tjhung, Nardini, and
  Cates}}]{MIPS10}
\bibinfo{author}{\bibfnamefont{E.}~\bibnamefont{Tjhung}},
  \bibinfo{author}{\bibfnamefont{C.}~\bibnamefont{Nardini}}, \bibnamefont{and}
  \bibinfo{author}{\bibfnamefont{M.~E.} \bibnamefont{Cates}},
  \bibinfo{journal}{Phys. Rev. X} \textbf{\bibinfo{volume}{8}},
  \bibinfo{pages}{031080} (\bibinfo{year}{2018}).

\bibitem[{\citenamefont{Speck et~al.}(2014)\citenamefont{Speck, Bialk\'e,
  Menzel, and L\"owen}}]{MIPS6}
\bibinfo{author}{\bibfnamefont{T.}~\bibnamefont{Speck}},
  \bibinfo{author}{\bibfnamefont{J.}~\bibnamefont{Bialk\'e}},
  \bibinfo{author}{\bibfnamefont{A.~M.} \bibnamefont{Menzel}},
  \bibnamefont{and} \bibinfo{author}{\bibfnamefont{H.}~\bibnamefont{L\"owen}},
  \bibinfo{journal}{Phys. Rev. Lett.} \textbf{\bibinfo{volume}{112}},
  \bibinfo{pages}{218304} (\bibinfo{year}{2014}).

\bibitem[{\citenamefont{Speck et~al.}(2015)\citenamefont{Speck, Menzel,
  Bialké, and L\"owen}}]{MIPS7}
\bibinfo{author}{\bibfnamefont{T.}~\bibnamefont{Speck}},
  \bibinfo{author}{\bibfnamefont{A.~M.} \bibnamefont{Menzel}},
  \bibinfo{author}{\bibfnamefont{J.}~\bibnamefont{Bialké}}, \bibnamefont{and}
  \bibinfo{author}{\bibfnamefont{H.}~\bibnamefont{L\"owen}},
  \bibinfo{journal}{J. Chem. Phys.} \textbf{\bibinfo{volume}{142}},
  \bibinfo{pages}{224109} (\bibinfo{year}{2015}).

\bibitem[{\citenamefont{Rapp et~al.}(2019)\citenamefont{Rapp, Bergmann, and
  Zimmermann}}]{MIPS8}
\bibinfo{author}{\bibfnamefont{L.}~\bibnamefont{Rapp}},
  \bibinfo{author}{\bibfnamefont{F.}~\bibnamefont{Bergmann}}, \bibnamefont{and}
  \bibinfo{author}{\bibfnamefont{W.}~\bibnamefont{Zimmermann}},
  \bibinfo{journal}{Eur. Phys. J. E} \textbf{\bibinfo{volume}{42}},
  \bibinfo{pages}{57} (\bibinfo{year}{2019}).

\bibitem[{\citenamefont{Landau and Lifshitz}(2013)}]{landau2013statistical}
\bibinfo{author}{\bibfnamefont{L.~D.} \bibnamefont{Landau}} \bibnamefont{and}
  \bibinfo{author}{\bibfnamefont{E.~M.} \bibnamefont{Lifshitz}},
  \emph{\bibinfo{title}{Statistical Physics: Volume 5}},
  vol.~\bibinfo{volume}{5} (\bibinfo{publisher}{Elsevier},
  \bibinfo{year}{2013}).

\bibitem[{\citenamefont{Wang et~al.}(2008)\citenamefont{Wang, Xu, and
  Wang}}]{flux1}
\bibinfo{author}{\bibfnamefont{J.}~\bibnamefont{Wang}},
  \bibinfo{author}{\bibfnamefont{L.}~\bibnamefont{Xu}}, \bibnamefont{and}
  \bibinfo{author}{\bibfnamefont{E.}~\bibnamefont{Wang}},
  \bibinfo{journal}{Proc. Natl. Acad. Sci. U. S. A.}
  \textbf{\bibinfo{volume}{105}}, \bibinfo{pages}{12271}
  (\bibinfo{year}{2008}).

\bibitem[{\citenamefont{Li and Wang}(2014)}]{flux2}
\bibinfo{author}{\bibfnamefont{C.}~\bibnamefont{Li}} \bibnamefont{and}
  \bibinfo{author}{\bibfnamefont{J.}~\bibnamefont{Wang}},
  \bibinfo{journal}{Proc. Natl. Acad. Sci. U. S. A.}
  \textbf{\bibinfo{volume}{111}}, \bibinfo{pages}{14130}
  (\bibinfo{year}{2014}).

\bibitem[{\citenamefont{Fang et~al.}(2019)\citenamefont{Fang, Kruse, Lu, and
  Wang}}]{flux3}
\bibinfo{author}{\bibfnamefont{X.}~\bibnamefont{Fang}},
  \bibinfo{author}{\bibfnamefont{K.}~\bibnamefont{Kruse}},
  \bibinfo{author}{\bibfnamefont{T.}~\bibnamefont{Lu}}, \bibnamefont{and}
  \bibinfo{author}{\bibfnamefont{J.}~\bibnamefont{Wang}},
  \bibinfo{journal}{Rev. Mod. Phys.} \textbf{\bibinfo{volume}{91}},
  \bibinfo{pages}{045004} (\bibinfo{year}{2019}).

\bibitem[{\citenamefont{Chu and Wang}(2020)}]{flux4}
\bibinfo{author}{\bibfnamefont{X.}~\bibnamefont{Chu}} \bibnamefont{and}
  \bibinfo{author}{\bibfnamefont{J.}~\bibnamefont{Wang}},
  \bibinfo{journal}{Appl. Phys. Rev.} \textbf{\bibinfo{volume}{7}},
  \bibinfo{pages}{031403} (\bibinfo{year}{2020}).

\bibitem[{\citenamefont{Wang}(2015)}]{flux5}
\bibinfo{author}{\bibfnamefont{J.}~\bibnamefont{Wang}}, \bibinfo{journal}{Adv.
  Phys.} \textbf{\bibinfo{volume}{64}}, \bibinfo{pages}{1}
  (\bibinfo{year}{2015}).

\bibitem[{\citenamefont{Fang and Wang}(2020)}]{flux7}
\bibinfo{author}{\bibfnamefont{X.}~\bibnamefont{Fang}} \bibnamefont{and}
  \bibinfo{author}{\bibfnamefont{J.}~\bibnamefont{Wang}},
  \bibinfo{journal}{Annu. Rev. Biophys.} \textbf{\bibinfo{volume}{49}}
  (\bibinfo{year}{2020}).

\bibitem[{\citenamefont{Seifert}(2012)}]{seifert2012stochastic}
\bibinfo{author}{\bibfnamefont{U.}~\bibnamefont{Seifert}},
  \bibinfo{journal}{Rep. Prog. Phys.} \textbf{\bibinfo{volume}{75}},
  \bibinfo{pages}{126001} (\bibinfo{year}{2012}).

\bibitem[{\citenamefont{Crochik and Tom{\'e}}(2005)}]{TP1}
\bibinfo{author}{\bibfnamefont{L.}~\bibnamefont{Crochik}} \bibnamefont{and}
  \bibinfo{author}{\bibfnamefont{T.}~\bibnamefont{Tom{\'e}}},
  \bibinfo{journal}{Phys. Rev. E} \textbf{\bibinfo{volume}{72}},
  \bibinfo{pages}{057103} (\bibinfo{year}{2005}).

\bibitem[{\citenamefont{Tom{\'e} and de~Oliveira}(2012)}]{TP2}
\bibinfo{author}{\bibfnamefont{T.}~\bibnamefont{Tom{\'e}}} \bibnamefont{and}
  \bibinfo{author}{\bibfnamefont{M.~J.} \bibnamefont{de~Oliveira}},
  \bibinfo{journal}{Phys. Rev. Lett.} \textbf{\bibinfo{volume}{108}},
  \bibinfo{pages}{020601} (\bibinfo{year}{2012}).

\bibitem[{\citenamefont{Noa et~al.}(2019)\citenamefont{Noa, Harunari,
  de~Oliveira, and Fiore}}]{TP3}
\bibinfo{author}{\bibfnamefont{C.~F.} \bibnamefont{Noa}},
  \bibinfo{author}{\bibfnamefont{P.~E.} \bibnamefont{Harunari}},
  \bibinfo{author}{\bibfnamefont{M.}~\bibnamefont{de~Oliveira}},
  \bibnamefont{and} \bibinfo{author}{\bibfnamefont{C.}~\bibnamefont{Fiore}},
  \bibinfo{journal}{Phys. Rev. E} \textbf{\bibinfo{volume}{100}},
  \bibinfo{pages}{012104} (\bibinfo{year}{2019}).

\bibitem[{\citenamefont{Seara et~al.}(2021)\citenamefont{Seara, Machta, and
  Murrell}}]{TP4}
\bibinfo{author}{\bibfnamefont{D.~S.} \bibnamefont{Seara}},
  \bibinfo{author}{\bibfnamefont{B.~B.} \bibnamefont{Machta}},
  \bibnamefont{and} \bibinfo{author}{\bibfnamefont{M.~P.}
  \bibnamefont{Murrell}}, \bibinfo{journal}{Nat. Commun.}
  \textbf{\bibinfo{volume}{12}}, \bibinfo{pages}{1} (\bibinfo{year}{2021}).

\bibitem[{\citenamefont{Xiao et~al.}(2008)\citenamefont{Xiao, Hou, and
  Xin}}]{TP5}
\bibinfo{author}{\bibfnamefont{T.~J.} \bibnamefont{Xiao}},
  \bibinfo{author}{\bibfnamefont{Z.}~\bibnamefont{Hou}}, \bibnamefont{and}
  \bibinfo{author}{\bibfnamefont{H.}~\bibnamefont{Xin}}, \bibinfo{journal}{J.
  Chem. Phys.} \textbf{\bibinfo{volume}{129}}, \bibinfo{pages}{114506}
  (\bibinfo{year}{2008}).

\bibitem[{\citenamefont{Shim et~al.}(2016)\citenamefont{Shim, Chun, and
  Noh}}]{TPA1}
\bibinfo{author}{\bibfnamefont{P.-S.} \bibnamefont{Shim}},
  \bibinfo{author}{\bibfnamefont{H.-M.} \bibnamefont{Chun}}, \bibnamefont{and}
  \bibinfo{author}{\bibfnamefont{J.~D.} \bibnamefont{Noh}},
  \bibinfo{journal}{Phys. Rev. E} \textbf{\bibinfo{volume}{93}},
  \bibinfo{pages}{012113} (\bibinfo{year}{2016}).

\bibitem[{\citenamefont{Cao et~al.}(2022)\citenamefont{Cao, Su, Jiang, and
  Hou}}]{TPA2}
\bibinfo{author}{\bibfnamefont{Z.}~\bibnamefont{Cao}},
  \bibinfo{author}{\bibfnamefont{J.}~\bibnamefont{Su}},
  \bibinfo{author}{\bibfnamefont{H.}~\bibnamefont{Jiang}}, \bibnamefont{and}
  \bibinfo{author}{\bibfnamefont{Z.}~\bibnamefont{Hou}},
  \bibinfo{journal}{Phys. Fluids} \textbf{\bibinfo{volume}{34}},
  \bibinfo{pages}{053310} (\bibinfo{year}{2022}).

\bibitem[{\citenamefont{Xu and Wang}(2020)}]{TPA3}
\bibinfo{author}{\bibfnamefont{L.}~\bibnamefont{Xu}} \bibnamefont{and}
  \bibinfo{author}{\bibfnamefont{J.}~\bibnamefont{Wang}}, \bibinfo{journal}{J.
  Phys. Chem. B} \textbf{\bibinfo{volume}{124}}, \bibinfo{pages}{2549}
  (\bibinfo{year}{2020}).

\bibitem[{\citenamefont{Fodor et~al.}(2016)\citenamefont{Fodor, Nardini, Cates,
  Tailleur, Visco, and Van~Wijland}}]{ep1}
\bibinfo{author}{\bibfnamefont{{\'E}.}~\bibnamefont{Fodor}},
  \bibinfo{author}{\bibfnamefont{C.}~\bibnamefont{Nardini}},
  \bibinfo{author}{\bibfnamefont{M.~E.} \bibnamefont{Cates}},
  \bibinfo{author}{\bibfnamefont{J.}~\bibnamefont{Tailleur}},
  \bibinfo{author}{\bibfnamefont{P.}~\bibnamefont{Visco}}, \bibnamefont{and}
  \bibinfo{author}{\bibfnamefont{F.}~\bibnamefont{Van~Wijland}},
  \bibinfo{journal}{Phys. Rev. Lett.} \textbf{\bibinfo{volume}{117}},
  \bibinfo{pages}{038103} (\bibinfo{year}{2016}).

\bibitem[{\citenamefont{Mandal et~al.}(2017)\citenamefont{Mandal, Klymko, and
  DeWeese}}]{ep2}
\bibinfo{author}{\bibfnamefont{D.}~\bibnamefont{Mandal}},
  \bibinfo{author}{\bibfnamefont{K.}~\bibnamefont{Klymko}}, \bibnamefont{and}
  \bibinfo{author}{\bibfnamefont{M.~R.} \bibnamefont{DeWeese}},
  \bibinfo{journal}{Phys. Rev. Lett.} \textbf{\bibinfo{volume}{119}},
  \bibinfo{pages}{258001} (\bibinfo{year}{2017}).

\bibitem[{\citenamefont{Dabelow et~al.}(2019)\citenamefont{Dabelow, Bo, and
  Eichhorn}}]{ep3}
\bibinfo{author}{\bibfnamefont{L.}~\bibnamefont{Dabelow}},
  \bibinfo{author}{\bibfnamefont{S.}~\bibnamefont{Bo}}, \bibnamefont{and}
  \bibinfo{author}{\bibfnamefont{R.}~\bibnamefont{Eichhorn}},
  \bibinfo{journal}{Phys. Rev. X} \textbf{\bibinfo{volume}{9}},
  \bibinfo{pages}{021009} (\bibinfo{year}{2019}).

\bibitem[{\citenamefont{Ganguly and Chaudhuri}(2013)}]{ep4}
\bibinfo{author}{\bibfnamefont{C.}~\bibnamefont{Ganguly}} \bibnamefont{and}
  \bibinfo{author}{\bibfnamefont{D.}~\bibnamefont{Chaudhuri}},
  \bibinfo{journal}{Phys. Rev. E} \textbf{\bibinfo{volume}{88}},
  \bibinfo{pages}{032102} (\bibinfo{year}{2013}).

\bibitem[{\citenamefont{Speck}(2016)}]{ep5}
\bibinfo{author}{\bibfnamefont{T.}~\bibnamefont{Speck}},
  \bibinfo{journal}{Europhys. Lett.} \textbf{\bibinfo{volume}{114}},
  \bibinfo{pages}{30006} (\bibinfo{year}{2016}).

\bibitem[{\citenamefont{Nardini et~al.}(2017)\citenamefont{Nardini, Fodor,
  Tjhung, Van~Wijland, Tailleur, and Cates}}]{ep6}
\bibinfo{author}{\bibfnamefont{C.}~\bibnamefont{Nardini}},
  \bibinfo{author}{\bibfnamefont{{\'E}.}~\bibnamefont{Fodor}},
  \bibinfo{author}{\bibfnamefont{E.}~\bibnamefont{Tjhung}},
  \bibinfo{author}{\bibfnamefont{F.}~\bibnamefont{Van~Wijland}},
  \bibinfo{author}{\bibfnamefont{J.}~\bibnamefont{Tailleur}}, \bibnamefont{and}
  \bibinfo{author}{\bibfnamefont{M.~E.} \bibnamefont{Cates}},
  \bibinfo{journal}{Phys. Rev. X} \textbf{\bibinfo{volume}{7}},
  \bibinfo{pages}{021007} (\bibinfo{year}{2017}).

\bibitem[{\citenamefont{Shankar and Marchetti}(2018)}]{ep7}
\bibinfo{author}{\bibfnamefont{S.}~\bibnamefont{Shankar}} \bibnamefont{and}
  \bibinfo{author}{\bibfnamefont{M.~C.} \bibnamefont{Marchetti}},
  \bibinfo{journal}{Phys. Rev. E} \textbf{\bibinfo{volume}{98}},
  \bibinfo{pages}{020604} (\bibinfo{year}{2018}).

\bibitem[{\citenamefont{Szamel}(2019)}]{ep8}
\bibinfo{author}{\bibfnamefont{G.}~\bibnamefont{Szamel}},
  \bibinfo{journal}{Phys. Rev. E} \textbf{\bibinfo{volume}{100}},
  \bibinfo{pages}{050603} (\bibinfo{year}{2019}).

\bibitem[{\citenamefont{Crosato et~al.}(2019)\citenamefont{Crosato, Prokopenko,
  and Spinney}}]{ep9}
\bibinfo{author}{\bibfnamefont{E.}~\bibnamefont{Crosato}},
  \bibinfo{author}{\bibfnamefont{M.}~\bibnamefont{Prokopenko}},
  \bibnamefont{and} \bibinfo{author}{\bibfnamefont{R.~E.}
  \bibnamefont{Spinney}}, \bibinfo{journal}{Phys. Rev. E}
  \textbf{\bibinfo{volume}{100}}, \bibinfo{pages}{042613}
  (\bibinfo{year}{2019}).

\bibitem[{\citenamefont{Cao et~al.}(2021)\citenamefont{Cao, Jiang, and
  Hou}}]{ep10}
\bibinfo{author}{\bibfnamefont{Z.}~\bibnamefont{Cao}},
  \bibinfo{author}{\bibfnamefont{H.}~\bibnamefont{Jiang}}, \bibnamefont{and}
  \bibinfo{author}{\bibfnamefont{Z.}~\bibnamefont{Hou}}, \bibinfo{journal}{J.
  Chem. Phys.} \textbf{\bibinfo{volume}{155}}, \bibinfo{pages}{234901}
  (\bibinfo{year}{2021}).

\bibitem[{\citenamefont{Tociu et~al.}(2019)\citenamefont{Tociu, Fodor, Nemoto,
  and Vaikuntanathan}}]{ep11}
\bibinfo{author}{\bibfnamefont{L.}~\bibnamefont{Tociu}},
  \bibinfo{author}{\bibfnamefont{{\'E}.}~\bibnamefont{Fodor}},
  \bibinfo{author}{\bibfnamefont{T.}~\bibnamefont{Nemoto}}, \bibnamefont{and}
  \bibinfo{author}{\bibfnamefont{S.}~\bibnamefont{Vaikuntanathan}},
  \bibinfo{journal}{Phys. Rev. X} \textbf{\bibinfo{volume}{9}},
  \bibinfo{pages}{041026} (\bibinfo{year}{2019}).

\bibitem[{\citenamefont{Nemoto et~al.}(2019)\citenamefont{Nemoto, Fodor, Cates,
  Jack, and Tailleur}}]{ep12}
\bibinfo{author}{\bibfnamefont{T.}~\bibnamefont{Nemoto}},
  \bibinfo{author}{\bibfnamefont{{\'E}.}~\bibnamefont{Fodor}},
  \bibinfo{author}{\bibfnamefont{M.~E.} \bibnamefont{Cates}},
  \bibinfo{author}{\bibfnamefont{R.~L.} \bibnamefont{Jack}}, \bibnamefont{and}
  \bibinfo{author}{\bibfnamefont{J.}~\bibnamefont{Tailleur}},
  \bibinfo{journal}{Phys. Rev. E} \textbf{\bibinfo{volume}{99}},
  \bibinfo{pages}{022605} (\bibinfo{year}{2019}).

\bibitem[{\citenamefont{Cagnetta et~al.}(2017)\citenamefont{Cagnetta, Corberi,
  Gonnella, and Suma}}]{ep13}
\bibinfo{author}{\bibfnamefont{F.}~\bibnamefont{Cagnetta}},
  \bibinfo{author}{\bibfnamefont{F.}~\bibnamefont{Corberi}},
  \bibinfo{author}{\bibfnamefont{G.}~\bibnamefont{Gonnella}}, \bibnamefont{and}
  \bibinfo{author}{\bibfnamefont{A.}~\bibnamefont{Suma}},
  \bibinfo{journal}{Phys. Rev. Lett.} \textbf{\bibinfo{volume}{119}},
  \bibinfo{pages}{158002} (\bibinfo{year}{2017}).

\bibitem[{\citenamefont{Guo et~al.}(2021)\citenamefont{Guo, Ro, Shih, Phan,
  Austin, Martiniani, Levine, and Chaikin}}]{ep14}
\bibinfo{author}{\bibfnamefont{B.}~\bibnamefont{Guo}},
  \bibinfo{author}{\bibfnamefont{S.}~\bibnamefont{Ro}},
  \bibinfo{author}{\bibfnamefont{A.}~\bibnamefont{Shih}},
  \bibinfo{author}{\bibfnamefont{T.~V.} \bibnamefont{Phan}},
  \bibinfo{author}{\bibfnamefont{R.~H.} \bibnamefont{Austin}},
  \bibinfo{author}{\bibfnamefont{S.}~\bibnamefont{Martiniani}},
  \bibinfo{author}{\bibfnamefont{D.}~\bibnamefont{Levine}}, \bibnamefont{and}
  \bibinfo{author}{\bibfnamefont{P.~M.} \bibnamefont{Chaikin}},
  \bibinfo{journal}{arXiv preprint arXiv:2105.12707}  (\bibinfo{year}{2021}).

\bibitem[{\citenamefont{Bowick et~al.}(2022)\citenamefont{Bowick, Fakhri,
  Marchetti, and Ramaswamy}}]{ep15}
\bibinfo{author}{\bibfnamefont{M.~J.} \bibnamefont{Bowick}},
  \bibinfo{author}{\bibfnamefont{N.}~\bibnamefont{Fakhri}},
  \bibinfo{author}{\bibfnamefont{M.~C.} \bibnamefont{Marchetti}},
  \bibnamefont{and}
  \bibinfo{author}{\bibfnamefont{S.}~\bibnamefont{Ramaswamy}},
  \bibinfo{journal}{Phys. Rev. X} \textbf{\bibinfo{volume}{12}},
  \bibinfo{pages}{010501} (\bibinfo{year}{2022}).

\end{thebibliography}

\end{document}